\documentclass[conference]{IEEEtran}
\usepackage{times}
\usepackage{listings}

\usepackage{cite}
\usepackage[misc]{ifsym}
\usepackage{amsmath,amssymb,amsfonts}
\usepackage[colorinlistoftodos]{todonotes}
\usepackage{bm}
\usepackage{listings}
\usepackage{graphicx}
\usepackage{textcomp}
\usepackage{graphicx}
\usepackage[numbers]{natbib}
\usepackage{multicol}
\usepackage{graphicx}
\usepackage{subfigure}
\usepackage[bookmarks=true]{hyperref}
\usepackage{mathtools}
\usepackage{amssymb,amsmath}
\usepackage[mathcal]{euscript}
\usepackage{xcolor}
\usepackage{booktabs}
\usepackage[symbol]{footmisc}
\usepackage{float}

\newfloat{Algorithm}{htbp}{lop}
\begin{document}

\title{A Novel Sentence Embedding Based Topic Detection Method For Micro-blog}

\author{\IEEEauthorblockN{Cong Wan, Shan Jiang$^{(\textrm{\Letter sancia1998@163.com})}$, Cuirong Wang$^{(\textrm{\Letter wangcr@mail.neuq.edu.cn})}$,\\ Cong Wang$^{(\textrm{\Letter congw@neuq.edu.cn})}$, Changming Xu, Xianxia Chen, Ying Yuan}
\IEEEauthorblockA{Northeastern University at Qinhuangdao, School of Computer and Communication Engineering}}

\maketitle

\begin{abstract}

Topic detection is a challenging task, especially without knowing the exact number of topics. In this paper, we present a novel approach based on neural network to detect topics in the micro-blogging dataset. We use an unsupervised neural sentence embedding model to map the blogs to an embedding space. Our model is a weighted power mean word embedding model, and the weights are calculated by attention mechanism. Experimental result shows our embedding method performs better than baselines in sentence clustering. In addition, we propose an improved clustering algorithm referred as relationship-aware DBSCAN (RADBSCAN). It can discover topics from a micro-blogging dataset, and the topic number depends on dataset character itself. Moreover, in order to solve the problem of parameters sensitive, we take blog forwarding relationship as a bridge of two independent clusters. Finally, we validate our approach on a dataset from sina micro-blog. The result shows that we can detect all the topics successfully and extract keywords in each topic.
\end{abstract}

\IEEEpeerreviewmaketitle


\section{Introduction}
In recent years, micro-blog platform has become an important place for people to share opinions, explore new events and disseminate information. In the micro-blog, various topics are forming, evolving and spreading every day. Topic detection from micro-blog is useful in many ways, such as natural disasters detection \cite{Ma2019Natural}, news recommendation \cite{Chuhan2019Neural}, community detection \cite{Zhou2017Analysis} and political analysis \cite{sasaki2017other}. Therefore, how to detect topics from a large microblog dataset has become an issue.
Topic detection is one of the key tasks in sentiment analysis. A large number of studies have been proposed in this field. For example, latent dirichlet allocation (LDA) \cite{Amoualian2017Topical} and various LDA-based models are widely used. They list some topics, such as food, sports and military, then calculate the probability that a document belongs to a topic. Neural network is another popular technology for topic detection in recently years. These networks take Bag-of-Words (BoW) as input and reconstruct the representation of documents. The output of neural network could be the final result or just used as part of the topic framework.
However, topic detection from micro-blogs is quite different and difficult. First of all, the micro blog is full of noise. Secondly, we don't know the number of topics. Finally, micro-blog has a network structure which can be used. 
In this work, we present a novel two-step approach to solve such issue. First, we design a sentence embedding neural model which maps a blog to embedding space. These sentence embeddings will be used for clustering, so we must make sure blogs with the same topic are mapped to nearby points. In our neural model, attention mechanism is adopted to make aspect more clear, and a power mean pooling layer is added to the model to make the embedding store different kinds of information. Then, we propose a clustering algorithm to discover clusters and the noise. In this algorithm, we improve DBSCAN \cite{ester1996a} by making use the network structure of blogs. We consider two points are in the same neighborhood when they are related in the micro blog network no matter how far away they are. The experiment result shows that our embedding method performs very better than other sentence embedding methods in the clustering task, and our clustering algorithm can accurately extract topics from micro-blog dataset and performs better than comparison algorithm.

\medskip
\section{Related Work}
\subsection{Topic Detection}
In the previous decades, much work has been done on topic detection and tracking. For the classical algorithm, the two main groups include text clustering and topic model. The relation of the two pages not only hinges on high repetition rate of words, but also depends on semantic relation. Because of Ambiguity and polysemy, topic model appeared. LSA (latent semantic analysis) is one of the major method to improve topic detection, which aimed at dealing with short text with little information we need and serious lack of semantics. The basic idea is to set text collection from sparse high-dimensional space to low-dimensional implicit semantic space. LDA is also a relatively effective topic model based on probability. The problem of LDA is sparsity. One way is to aggregate multiple short texts into one long text with the assistance of external information. The other is to extract information of a chosen long text into a priori in some form first, but it is a trouble to look for a suitable long text. Thus it is difficult to effectively solve the problems in topic detection with a single method. Amoualian et al. \cite{Amoualian2017Topical} present an LDA-based model that generates topically coherent segments within documents by jointly segmenting documents and assigning topics to their words. Some scholars have used other methods to replace the above topic model, and have also achieved good results. Gallagher et al. \cite{Gallagher2017Anchored} proposed Topic Modeling with Minimal Domain Knowledge by Correlation Explanation to produce rich topics, optimizing Correlation explanation framework of sparse binary data. Yang et al. \cite{Yang2016A} make use of the rich link structure of the document network to improve topic coherence. Hida et al. \cite{Hida2018Dynamic} proposed a dynamic and static model, considering both the dynamic structure of temporal topic evolution and the static structure of each topic hierarchy. 

\subsection{Sentence Embedding}
For sentence embedding, if the order of words in a sentence is ignored, it is considered a bag of words. But BOW (bag of words) is not effective for tasks that are sensitive to the order of words, such as sentiment analysis. While RNN (Recurrent Neural Network) overcomes shortcomings of BOW, it is often combined with supervised tasks, and it lacks scalability or transferability. Skip-thought is a typical example of learning unsupervised sentence embeddings. It relies on the idea of skip-gram. This model reconstructs the sentences surrounding the current sentence and consists of an RNN-based encoder-decoder. There have been some new developments in the last two years. Arora et al. \cite{arora2017a} claim to find a simple but tough-to-beat basline model. Each sentence is expressed as a weighted average of the embedding of the contained words, then the sentences are put together to find the largest major axis; and finally the largest major axis is removed from each sentence.  Alexis Conneau et al. \cite{Conneau2017Supervised} use SNLI (Stanford Natural Language Inference Corpus), train the model in advance, and then transfer learning. In SNLI, each pair of sentences is marked as one of three categories: entailment, contradiction, and neutral. The selected model BiLSTM-Max works best and outperforms the other models. Logeswaran et al. \cite{Logeswaran2018An} propose a simple and efficient framework named QT (Quick Thoughts) based Skip-thought. The difference between QT and skip-thought is that QT is directly output to a classifier after encoding, and judges the adjacent sentence. The advantage is that the speed of operation is greatly improved.

\subsection{Attention Mechanism}
Attention mechanism has been widely used in Deep Learning such as Natural Language Processing, Image Processing, and speech signal processing in recent years, and many variations have also appeared.  As the name suggests, the attention model draws on the human attention mechanism, the focus of human attention. The attention model improves the efficiency and accuracy of deep learning tasks by investing more "attention" into high-value information that requires focus. The more successful application of the attention mechanism is neural machine translation. The attention mechanism is introduced in the "encoding-decoding" mode of traditional neural machine translation, so that the fixed interlingua semantics are replaced with the semantics changing according to the generated words as the input of the decoder, which greatly improves the accuracy of translation. After Mnih et al. \cite{Mnih2014Recurrent} used the attention mechanism to classify images on the RNN model, the attention mechanism began to be widely used by scholars in various fields. Bahdanau et al. \cite{bahdanau2015neural} used a similar attention mechanism on the encoding-decoding framework when performing machine translation tasks which made the sequence adjustment and phrase translation process performing simultaneously, so that the model can automatically select the relevant part of the original sentence as the input of decoder when predicting the next word. Lmthang et al. \cite{luong2015effective} mentioned two attention mechanisms: one is global mechanism and the other is local mechanism. They showed how attention could be extended in RNN, which promoted the subsequent application of attention-based models in NLP. Yin et al. \cite{Yin2015ABCNN} proposed three schemes based on attention. Considering the mutual influence of two sentences under the same task background, the influence is incorporated into CNN. This is the early exploration of attention in CNN. Vaswani et al. \cite{vaswani2017attention} get rid of the traditional encoding-decoding model combined with CNN or RNN in the model with only attention mechanism instead, and used Multi-headed self-attention in the encoder and decoder. Tan et al. \cite{tan2018deep} proposed to label semantic roles with deep attention networks. Self-attention can be seen as a special case of general attention. In NLP tasks, self-attention can also be used as a layer together with RNN.

\subsection{ Social network structure analysis}
Complex network is a form of data. In real life, analyzing a complex system can be modeled as a complex network. The nodes in a complex network will exhibit cluster characteristics. Such as transportation networks, aviation networks, computer networks, and social networks that are closest to human activities. As the so-called things gather together, people divide into groups. Microblogs, Facebook, and Renren all belong to social networks. These social networks provide participants with a platform where people can show themselves, develop hobbies, and use these hobbies to socialize. A large number of active users provide rich data and scenarios for researching information dissemination, community division, social network recommendation, topic detection and evolution. As far as the recommendation system is concerned,  Hao et al. \cite{Hao2008SoRec} proposed a factor analysis method based on probability matrix factorization, which solves the data by simultaneously using the user’s social network information and score records Problems of sparseness and poor prediction accuracy. It is different from the traditional assumption that the system is independent of the user. Zhou et al. \cite{Zhou2018Academic} have proposed an academic influence perception and multidimensional network analysis method, which has made certain contributions to meet the limitations of heterogeneous network analysis. For example, most recommendation systems only focus on a single relationship or two or three relationships when constructing a network model, and the result can only be regarded as some hierarchical network or multi-layer graph. Jamali et al. \cite{jamali2010a} incorporated the trust propagation mechanism into the model, and proposed a matrix-based social network recommendation method using matrix decomposition technology. For users, sharing daily and some personal feelings on social networks has become a way of life. Zhou et al. \cite{Xiaokang2019Multi} proposed a learning model based on integrated deep neural network , focusing on the behavioral influence analysis based on heterogeneous health data generated in social media environments. 

\medskip
\section{Method}
\begin{figure}
    \centering
    \includegraphics[width=.92\linewidth]{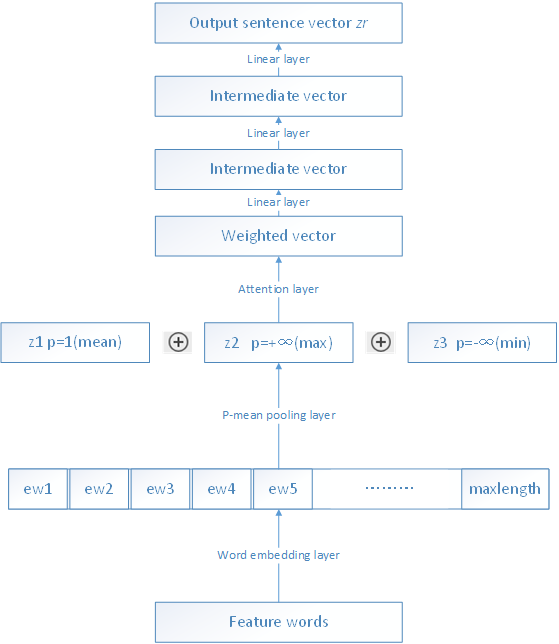}
    \caption{model}
    \label{fig:model}
\end{figure}

In this section, we will introduce our approach in detail.\ref{fig:model} First, a power-mean and attention based neural model(PANM) is proposed to generate vector from micro-blog, furthermore, this model is unsupervised. Secondly, a novel clustering algorithm referred as RADBSCAN is proposed to detect topics.

\medskip
\subsection{Neural Model}
The goal of our PANM model is to learn sentence embeddings. The structure of PANM is shown in Figure 1. Firstly, the micro-blogs are segmented and some feature words are selected as the input of our model. Stop words, numbers, punctuations and useless words such as“@someone”are removed. Each feature word is associated to a feature vector $ e_w^{} \in R_{}^d\ $ , where d is the dimension of word embedding. Then we construct a sentence embedding z from the vector   by a power mean based method\cite{R2018Concatenated}. Power mean generalized the idea of generate sentence embedding by average word embedding, it can be described as (1), where p defines mean type such as such as the arithmetic mean (p = 1), the geometric mean (p = 0), and the harmonic mean (p = -1). 
\begin{equation}
z = (\sum\limits_{i = 1}^n {e_{wi}^p} )_{}^{1/p}
\end{equation}
In our work, we weighted the power mean method as (2)(3), so some more important keyword can have greater impact on the sentence position in embedding space, which could make sentences with same topic closer.
\begin{equation}
z = (\sum\limits_{i = 1}^n {a_i^{}e_{wi}^p} )_{}^{1/p}
\end{equation}  \begin{equation}
\sum\limits_{i = 1}^n {a_i^{} = 1} 
\end{equation}
Moreover, in order to get a richer summary statistics of the sentence, we select some different p values and concatenate the sentence embeddings together. Thus we get a new embedding z.
\begin{equation}
z = (\sum\limits_{i = 1}^n {a_{1i}^{}e_{wi}^{p_1^{}}} )_{}^{1/p_1^{}} \oplus (\sum\limits_{i = 1}^n {a_{2i}^{}e_{wi}^{p_2^{}}} )_{}^{1/p_2^{}} \oplus (\sum\limits_{i = 1}^n {a_{3i}^{}e_{wi}^{p_3^{}}} )_{}^{1/p_3^{}}
\end{equation}
The word weight a is computed by an attention-based method.
\begin{equation}
y = \frac{{(\sum\limits_{i = 1}^n {e_{wi}^p} )_{}^{1/p}}}{n}
\end{equation}
\begin{equation}
d_i^{} = e_{wi}^T \cdot M \cdot y
\end{equation}
\begin{equation}
[a_i^{} = \frac{{e_{}^{d_i^{}}}}{{\sum\limits_{i = 1}^n {e_{}^{d_i^{}}} }}]
\end{equation}

Finally, we reconstruct sentence embedding zr through three linear layers. This process can be described as the following formula (8), where  is the activation function ReLU.
\begin{equation}
zr = g(g(g(z_{}^T \cdot M_1^{}) \cdot M_2^{}) \cdot M_3^{})
\end{equation}

So, the goal of training PANM is to learn the parameter matrices, M, M1, M2 and M3.

After the description of our model structure, we now introduce our loss function L which is based on hingle loss\cite{He2018UNSUPERVISED, Socher2013Grounded}. For training, we use L to make sure the following target: 1) there is a high inner product between zr and z, 2) there are low inner products between zr and negative samples. The description of our loss function is as (9), where is a negative sample.
\begin{equation}
L = \sum\limits_{i = 1}^m {\max (0,1 - z_{}^Tzr + zr_{}^Ts_i^{})} 
\end{equation}

The negative samples are randomly selected from the micro-blog dataset. when we get a random blog, we associate its feature words to word embeddings as the first layer of PANM.
Then the negative sample s is generated by an unweighted power mean concatenation method as (10).
\begin{equation}
s = (\sum\limits_{i = 1}^n {e_{wi}^{p_1^{}}} )_{}^{1/p_1^{}} \oplus (\sum\limits_{i = 1}^n {e_{wi}^{p_2^{}}} )_{}^{1/p_2^{}} \oplus (\sum\limits_{i = 1}^n {e_{wi}^{p_3^{}}} )_{}^{1/p_3^{}}
\end{equation}

\medskip
\subsection{relationship-aware DBSCAN (RADBSCAN)}
\begin{figure}
\centering
\subfigure[]{
\begin{minipage}{0.5\textwidth}
    \includegraphics[width=.92\linewidth]{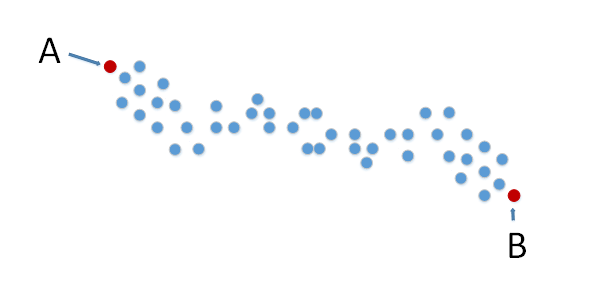}
    \label{fig:2a}
\end{minipage}
}
\\
\subfigure[]{
    \begin{minipage}{0.5\textwidth}
    \includegraphics[width=.92\linewidth]{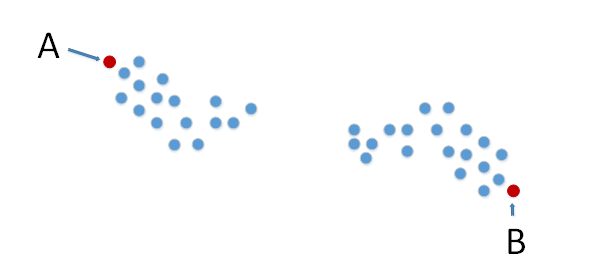}
    \label{fig:2b}
\end{minipage}
}
\\
\subfigure[]{
    \begin{minipage}{0.5\textwidth}
    \includegraphics[width=.92\linewidth]{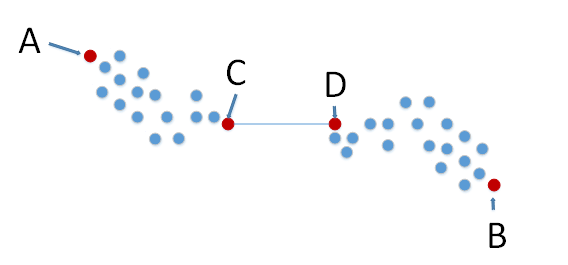}
    \label{fig:2c}
\end{minipage}
}
\caption{}
\end{figure}

In this subsection, we describe our algorithm relationship-aware density based spatial clustering of applications with noise(RADBSCAN) that make use of forwarding relationship between blogs. The main idea of our algorithm is to expand the definition of density-reachable[DBSCANkdd-96] which is the core concept of DBSCAN. If there is a path from point A to point B, which is surrounded by points whose density exceeds a certain threshold, then point A and point B are density-reachable, such as Figure \ref{fig:2a}. In contrast, point A and point B are density-unreachable in Figure \ref{fig:2b} because the path is broken. However, in our opinion, if there is an exact connection between point C and point D, then they can rebuild the path. So A and B are still density-reachable in Figure \ref{fig:2c}. Forwarding relationship between blogs can be this exact connection.

Based on the above viewpoint, RADBSCAN algorithm is proposed. In the perspective of clustering algorithms, sentence vectors are regarded as points in embedding space. So in the description our algorithms, the notion of point represents a sentence. In addition, there are two other important concepts as follows:
Definition 1: neighborhood-of-a-point (short for eps) is a threshold defining the range of neighbors. If the distance of two points is less than eps, then they are in the same neighborhood.
Definition 2: minimum-number-of-points (short for MinPts) is a threshold defining the minimum number of points in the same neighborhood.
In algorithm\ref{algorithm:algorithm1}, D is the whole points dataset, each points in D has a label and a state. The label indicates which cluster it belongs to, and all points have no label in the beginning. Each point in D belongs to one of the following three states: visited, noise or undefined. Visited point already has cluster label; noise point supposed to be a noise without label, but there's still a chance to get the label; undefined is initial state.

\begin{Algorithm}[t]
\begin{lstlisting} [frame=single]  
Input: D , eps, MinPts
Output: clusters with different labels C
   C = 0
   foreach point P in dataset D {
      if P is visited or P is NOISE
        continue next point
      [NeighborPts, RelatedPts] = regionQuery(P, eps)
      If sizeof(NeighborPts) < MinPts
        mark P as NOISE
     else {
        C = C+1
        mark P as visited
        NeighborPts= NeighborPts joined with RelatedPts
        expandCluster(P, NeighborPts, C, eps, MinPts)
      }
   }
\end{lstlisting}
\caption{RADBSCAN}
\label{algorithm:algorithm1}
\end{Algorithm}

\medskip
\begin{Algorithm}[t]
\begin{lstlisting} [frame=single]
Input: P, eps
NeighborPts= all points within P's eps-neighborhood (including P)
RelatedPts=all points having forwarding relationship 
Return [NeighborPts, RelatedPts]
\end{lstlisting}
\caption{regionQuery}
\label{}
\end{Algorithm}

\medskip
\begin{Algorithm}[t]
\begin{lstlisting} [frame=single]
Input: P NeighborPts, C, eps, MinPts
Output: Cluster C with all of its members
add P to cluster C
   foreach point P' in NeighborPts {
      if P' is not visited {
        mark P' as visited
        [NeighborPts', RelatedPts']= regionQuery(P', eps)
        
        if sizeof(NeighborPts') >= MinPts
           NeighborPts = NeighborPts joined with NeighborPts'
        NeighborPts= NeighborPts joined with RelatedPts'
      }
      If P' is not yet member of any cluster
        add P' to cluster C
\end{lstlisting}
\caption{expandCluster}
\label{}
\end{Algorithm}

\medskip
\section{Performance Evaluation}
In this section, we evaluate the performance our method including both sentence embedding model PANM and clustering algorithm RADBSCAN.

\subsection{Dataset}
Our evaluation is based on a real micro-blog dataset which is crawled from www.sina.com.
There are 15000 blogs in the dataset including five topics: cellphone, basketball, house prices, civil servants and smog. In order to test the performance of our method in different topic numbers and micro-blog numbers, we randomly extract topics from the whole dataset and form several subsets. Their identifiers are as Table \ref{table:I}, in which Meizu is a popular Chinese mobile phone brand and Rocket is a NBA team. 

\begin{table}[t]
\setlength{\tabcolsep}{3.5mm}
\begin{tabular}{lll}
\hline
Identity  & Topics                                                                                                     & Blog Number \\ \hline
Dataset A & \begin{tabular}[c]{@{}l@{}}Meizu, Rocket, house prices, \\ civil servants and smog\end{tabular} & 15000       \\ 
Dataset B & \begin{tabular}[c]{@{}l@{}}Meizu, Rocket, house prices,\\ civil servants\end{tabular}          & 12000       \\ 
Dataset C & \begin{tabular}[c]{@{}l@{}}Meizu, Rocket,  house prices\end{tabular}                          & 9000        \\ \hline
\end{tabular}
\medskip
\caption{ \textbf{DATASET identifiers}}
\label{table:I}
\end{table}

\subsection{effect evaluation of sentence embedding}
In this work, we propose a sentence embedding model PANM, through which sentence vectors are generated, and then these vectors are used for short text clustering. Our model is trained on a PC with 8 GB memory. The software environment is python3.6, pycharm and numpy. The word embedding dimension is 100, epoch is 10, negative sample set size is 20, optimizer is adam and learning rate is 0.001.
In order to evaluate PANM, we choose the following sentence embedding methods as baselines, and then clustering with K-mean and EM. The result of clustering can be used to evaluate the model. Baselines are as follow:

\begin{itemize}
    \item \textbf{Power-mean \cite{R2018Concatenated}:} This model is similar to ours, which adopts a variety of pooling methods and joins them. But we have an obvious difference. We use the attention model to give weight to every word in a sentence, so that the most important word has the highest weight.
    \item \textbf{TF-IDF-SVD:} The feature of a sentence can be represented by a sparse vector of TF-IDF, and then the vector is reduced by SVD.
    \item \textbf{SIF \cite{arora2017a}:} It is the most common baseline in sentence embedding.
    \item \textbf{Simple Word Averaging \cite{wieting2015towards}:}SWA is a simple but well-performed method. It is the inspiration of many methods.
    \item \textbf{Keywords- averaging :} We can extract three most important keywords according to word weights from three pooling method (mean, max, min) in PANM. KM is a simple average of these three keywords' embedding.
\end{itemize}

\medskip
We choose some metrics for performance evaluation. Before we introduce these metrics, let's define some symbols. Give a set of N samples  $  S = \{ o_1^{},o_2^{}...o_N^{}\}$ , let  $ C = \{ c_1^{},c_2^{}...c_k^{}\} $  be the true partition of the sample set, and  $ \Omega  = \{ \omega _1^{},\omega _2^{}...\omega _k^{}\} $ be the calculated partition of sample set. 

\begin{itemize}
    \item \textbf{NMI (Normalized Mutual Information)} is a measure of the similarity between two labels of the same sample set. NMI can be calculated as (11). The value of NMI range from 0 to 1. The larger the value is, the better the classification effect is.
\begin{equation}
NMI = \frac{{2\sum\limits_{i = 1}^k {\sum\limits_{j = 1}^k {\frac{{|\omega _i^{} \cap c_j^{}|}}{N}\log \frac{{N|\omega _i^{} \cap c_j^{}|}}{{|\omega _i^{}||c_j^{}|}}} } }}{{ - \sum\limits_{i = 1}^k {\frac{{|\omega _i^{}|}}{N}\log \frac{{|\omega _i^{}|}}{N} - \sum\limits_{j = 1}^k {\frac{{|c_j^{}|}}{N}\log \frac{{|c_j^{}|}}{N}} } }}
\end{equation}
    \item \textbf{RI (Rand Index)} is used for measure of the similarity between two data clusterings. RI is similar to accuracy, but it is more widely used in measurement of clustering algorithm. JC can be calculated as (12), in which  $ a = |S_{}^*| $  where  $ S_{}^* = \{ (o_i^{},o_j^{})|o_i^{},o_j^{} \in \omega _l^{},o_i^{},o_j^{} \in c_g^{}\} $  and  $ b = |S_{}^*| $   where  $ S_{}^* = \{ (o_i^{},o_j^{})|o_i^{} \notin \omega _l^{},o_j^{} \in \omega _l^{},o_i^{} \notin c_g^{},o_j^{} \in c_g^{}\}  $.
\begin{equation}
RI = \frac{{a + b}}{{2N(N - 1)}}
\end{equation}
    \item \textbf{JC (Jaccard Coefficient)} is a statistic used for gauging the similarity and diversity of sample sets. NMI can be calculated as (13), in which in which  $ a = |S_{}^*| $  where    $ S_{}^* = \{ (o_i^{},o_j^{})|o_i^{},o_j^{} \in \omega _l^{},o_i^{} \notin c_g^{},o_j^{} \in c_g^{}\} $ .
\begin{equation}
JC{\rm{ = }}\frac{a}{{a + b + c}}
\end{equation}
    \item \textbf{FMI (Fowlkes and Mallows Index)} is an external evaluation method that is used to determine the similarity between two clusterings. It can be calculated as (14).
\begin{equation}
FMI = \sqrt {\frac{a}{{a + b}} \bullet \frac{a}{{a + c}}}
\end{equation}
    \item \textbf{Precision} can be calculated as (15).
\begin{equation}
\Pr ecision = \sum\limits_{i = 1}^k {\frac{{|\omega _i^{}|}}{N}} \max \{ \frac{{|c_j^{} \cap \omega _i^{}|}}{{|\omega _i^{}|}},c_j^{} \in C\} 
\end{equation}
\end{itemize}
Then we clustering those sentence embeddings with K-mean and EM algorithms. The results are as follows. Finally, we find that our results are the best, especially the value of NMI. it proves that our method can better extract the feature of a micro-blog in Chinese text clustering. 

\begin{table*}[]
\renewcommand\arraystretch{1.5}
	\centering
	\setlength{\tabcolsep}{4mm}{
\begin{tabular}{lllllllllll}
\hline
\textbf{}                                                        & K-mean &       &       &       &                               & EM    &       &       &       &           \\
                                                                 & NMI    & JC    & RI    & FMI   & \multicolumn{1}{c}{Rrecision} & NMI   & JC    & RI    & FMI   & Rrecision \\ \hline
PANM                                                             & 0.628  & 0.391 & 0.782 & 0.574 & 0.726                         & 0.609 & 0.387 & 0.789 & 0.579 & 0.67      \\
Power-mean                                                       & 0.619  & 0.406 & 0.785 & 0.532 & 0.759                         & 0.601 & 0.365 & 0.741 & 0.538 & 0.669     \\
\begin{tabular}[c]{@{}l@{}}TF-IDF-\\ SVD\end{tabular}            & 0.57   & 0.327 & 0.742 & 0.505 & 0.703                         & 0.456 & 0.34  & 0.781 & 0.51  & 0.632     \\
SIF                                                              & 0.576  & 0.347 & 0.747 & 0.53  & 0.691                         & 0.421 & 0.304 & 0.753 & 0.471 & 0.578     \\
\begin{tabular}[c]{@{}l@{}}Simple Word \\ Averaging\end{tabular} & 0.578  & 0.350 & 0.751 & 0.532 & 0.696                         & 0.446 & 0.324 & 0.769 & 0.493 & 0.616     \\
\begin{tabular}[c]{@{}l@{}}Keywords-\\ averaging\end{tabular}    & 0.624  & 0.402 & 0.726 & 0.517 & 0.714                         & 0.427 & 0.272 & 0.645 & 0.465 & 0.558     \\ \hline
\end{tabular}
}
\medskip
\caption{ Dataset A}
\end{table*}
\begin{table*}[]
\renewcommand\arraystretch{1.5}
	\centering
	\setlength{\tabcolsep}{4mm}{
\begin{tabular}{lllllllllll}
\hline
\textbf{}                                                        & K-mean &       &       &       &                               & EM    &       &       &       &           \\
                                                                 & NMI    & JC    & RI    & FMI   & \multicolumn{1}{c}{Rrecision} & NMI   & JC    & RI    & FMI   & Rrecision \\ \hline
PANM                                                             & 0.616  & 0.441 & 0.780 & 0.616 & 0.777                         & 0.643 & 0.487 & 0.803 & 0.660 & 0.778     \\
Power-mean                                                       & 0.609  & 0.417 & 0.777 & 0.563 & 0.770                         & 0.633 & 0.406 & 0.772 & 0.656 & 0.772     \\
\begin{tabular}[c]{@{}l@{}}TF-IDF-\\ SVD\end{tabular}            & 0.513  & 0.337 & 0.685 & 0.516 & 0.684                         & 0.420 & 0.372 & 0.733 & 0.548 & 0.651     \\
SIF                                                              & 0.524  & 0.357 & 0.692 & 0.540 & 0.672                         & 0.446 & 0.348 & 0.746 & 0.517 & 0.629     \\
\begin{tabular}[c]{@{}l@{}}Simple Word \\ Averaging\end{tabular} & 0.534  & 0.365 & 0.701 & 0.548 & 0.682                         & 0.426 & 0.325 & 0.729 & 0.491 & 0.627     \\
\begin{tabular}[c]{@{}l@{}}Keywords-\\ averaging\end{tabular}    & 0.607  & 0.437 & 0.772 & 0.631 & 0.776                         & 0.425 & 0.299 & 0.625 & 0.480 & 0.612     \\ \hline
\end{tabular}
}
\medskip
\caption{ Dataset B}
\end{table*}
\begin{table*}[]
\renewcommand\arraystretch{1.5}
	\centering
	\setlength{\tabcolsep}{4mm}{
\begin{tabular}{lllllllllll}
\hline
\textbf{}                                                        & K-mean &       &       &       &                               & EM    &       &       &       &           \\
                                                                 & NMI    & JC    & RI    & FMI   & \multicolumn{1}{c}{Rrecision} & NMI   & JC    & RI    & FMI   & Rrecision \\ \hline
PANM                                                             & 0.509  & 0.447 & 0.713 & 0.622 & 0.757                         & 0.454 & 0.431 & 0.714 & 0.605 & 0.744     \\
Power-mean                                                       & 0.485  & 0.423 & 0.709 & 0.768 & 0.717                         & 0.421 & 0.344 & 0.690 & 0.606 & 0.730     \\
\begin{tabular}[c]{@{}l@{}}TF-IDF-\\ SVD\end{tabular}            & 0.411  & 0.366 & 0.611 & 0.548 & 0.664                         & 0.421 & 0.43  & 0.712 & 0.61  & 0.733     \\
SIF                                                              & 0.501  & 0.403 & 0.700 & 0.612 & 0.704                         & 0.322 & 0.345 & 0.656 & 0.514 & 0.741     \\
\begin{tabular}[c]{@{}l@{}}Simple Word \\ Averaging\end{tabular} & 0.506  & 0.408 & 0.704 & 0.616 & 0.709                         & 0.324 & 0.332 & 0.641 & 0.5   & 0.647     \\
\begin{tabular}[c]{@{}l@{}}Keywords-\\ averaging\end{tabular}    & 0.497  & 0.407 & 0.713 & 0.611 & 0.737                         & 0.277 & 0.326 & 0.590 & 0.499 & 0.616     \\ \hline
\end{tabular}
}
\medskip
\caption{ Dataset C}
\end{table*}

\medskip
\subsection{Experimental analysis of RADBSCAN}
\begin{figure}[t]
    \centering
    \includegraphics[width=.92\linewidth]{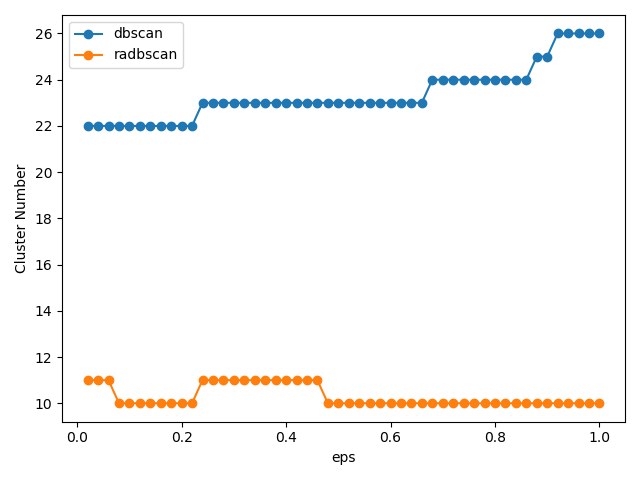}
    \caption{the number of clusters of DBSCAN and RADBSCAN}
    \label{fig:3}
\end{figure}
\begin{figure}[t]
    \centering
    \includegraphics[width=.92\linewidth]{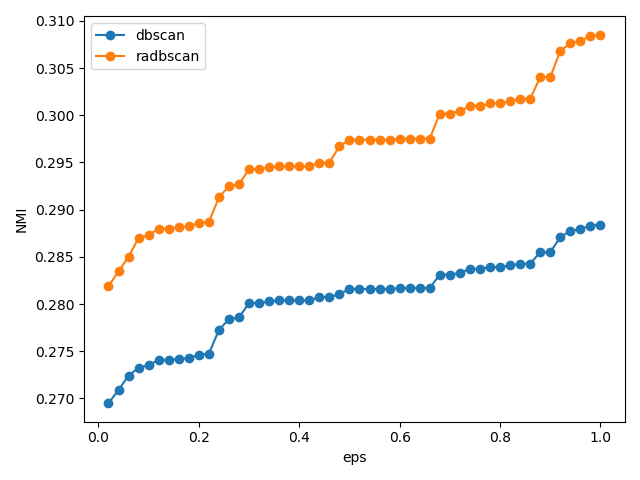}
    \caption{the NMI of DBSCAN and RADBSCAN }
    \label{fig:4}
\end{figure}
In order to show the effectiveness of the proposed algorithm, performance evaluation based on the foregoing work is conducted. The goal of this work is pick out topics worthy of attention from a large number of noises. DBSCAN is designed for this purpose, however, there are some disadvantages. In this section, we will show that our RADBSCAN has enhanced DBSCAN and avoids two disadvantages. First, DBSCAN is parameter sensitive. The value of eps has great influence on clustering results. Secondly, high dimension reduces DBSCAN performance.
Figure \ref{fig:3} shows the number of clusters in dataset A obtained by two DBSCAN and RADBSCAN. In RADBSCAN, number of clusters is very stable, and the influence of eps is very small. This shows that our algorithm is not as eps sensitive as DBSCAN. In addition, our number of clusters is very close to the number of real clusters which is 5. Figure \ref{fig:4} shows the NMI of two algorithms. It shows that our clustering results are closer to the real clusters.
Table \ref{table:V} shows the keywords extracted from these clusters. The topics of these clusters are very clear. The first, second, third and seventh clusters are about Meizu cellphone, rocket team, civil servants and house price respectively. The fourth cluster has three keywords: air-blast, square and army green. This cluster describes a remarkable event, the appearance of army green air-blast machines in the square. Smog was a focus for public discontent in China, so the first appearance of air-blast sprayer has attracted great attention. It is believed that this invention can control smog. The news is still visible on Chinese websites. The sixth cluster is about fog and wind. It's hard for ordinary people to distinguish between natural fog and pollution smog. But our algorithm separates these two topics. From the clustering results, our algorithm not only finds all topics, but also has better effect than manual label.

\begin{table}[]
\renewcommand\arraystretch{1.5}
	\centering
	\setlength{\tabcolsep}{10mm}{
\begin{tabular}{ll}
\hline
Cluster & Keywords                    \\ \hline
1       & Meizu                       \\
2       & Rockets                     \\
3       & Civil servants              \\
4       & Air-blast; Square; Army green \\
5       & Smog                        \\
6       & Wind; Fog                   \\
7       & House price                 \\ \hline
\end{tabular}
}
\caption{keywords extracted from these clusters}
\label{table:V}
\end{table}

\section{Conclusions}
Our work includes: (1) we have presented PANM, a neural attention model for sentence embedding. In contrast to comparison methods, PANM performs better in sentence clustering. (2) we have proposed RADBSCAN, an improved density based approach for micro-blog sentence clustering. It benefits from the forwarding relationships between blogs and performs better than DBSCAN. Our work provides a simple method of finding topic from micro-blog. We believe that our work could help people make better use of micro-blog data.

\section{Acknowledgements}
This work was supported by the National Natural Science Foundation of China under Grant Nos. 61702089, the Basic scientific research operating found of central universities under Grant No. N182304021, the Scientific research plan for institutions of higher learning of Hebei province under Grant No. ZD2019306, and the China Scholarship Council under Grant No. 201908130244.

\bibliographystyle{IEEEtran}



\end{document}